\documentclass{article}

\usepackage{arxiv}

\usepackage[utf8]{inputenc} 
\usepackage[T1]{fontenc}    
\usepackage{hyperref}       
\usepackage{url}            
\usepackage{booktabs}       
\usepackage{amsfonts}       
\usepackage{nicefrac}       
\usepackage{microtype}      
\usepackage{lipsum}
\usepackage{graphicx}
\graphicspath{ {./images/} }
\usepackage{algorithm}%
\usepackage{algpseudocode}
\usepackage{subcaption}
\usepackage{xcolor}
\usepackage{array}    
\usepackage{tabularx} 
\usepackage{amsmath}
\usepackage{orcidlink}
\title{A Multi-Model Approach Using XAI and Anomaly Detection to Predict Asteroid Hazards}

\author{
\orcidlink{0009-0000-0353-9915} Amit Kumar Mondal \\
  Computer Science and Engineering\\
 Bengal College of Engg. and Tech.\\India,WB, 713212 \\
  \texttt{mamit5620@gmail.com} \\
  \And
 Nafisha Aslam \\
  Computer Science and Engineering\\
  Bengal College of Engg. and Tech.\\India,WB, 713212 \\
  \texttt{nafishaaslam2@gmail.com} \\ \\
   \And
  \orcidlink{0000-0001-8057-6963} Prasenjit Maji \\
  Computer Science and Design\\
  Dr. B. C. Roy Engineering College\\India,WB, 713206 \\
  \texttt{maji.katm@gmail.com} \\
   \And
 \orcidlink{0000-0002-9403-4724} Hemanta Kumar Mondal \\
  Electronics and Communication Engg.\\
  National Institute of Technology\\
  Durgapur,India, WB, 713209 \\
  \texttt{hkmondal.ece@nitdgp.ac.in} \\
}

\begin{document}
\maketitle
\begin{abstract}
The potential for catastrophic collision makes near-Earth asteroids (NEAs) a serious concern. Planetary defense depends on accurately classifying potentially hazardous asteroids (PHAs), however the complexity of the data hampers conventional techniques. This work offers a sophisticated method for accurately predicting hazards by combining machine learning, deep learning, explainable AI (XAI), and anomaly detection. Our approach extracts essential parameters like size, velocity, and trajectory from historical and real-time asteroid data. A hybrid algorithm improves prediction accuracy by combining several cutting-edge models. A forecasting module predicts future asteroid behavior, and Monte Carlo simulations evaluate the likelihood of collisions. Timely mitigation is made possible by a real-time alarm system that notifies worldwide monitoring stations. This technique enhances planetary defense efforts by combining real-time alarms with sophisticated predictive modeling.
\end{abstract}


\section{Introduction}
Asteroids are tiny rocky objects that orbit the Sun; some could collide with Earth and cause severe damage. Impact risk assessment and mitigation depend heavily on the ability to predict asteroid risks. Machine learning has become crucial for effectively processing and interpreting the growing amount of astronomical data \cite{Singh2021}. Recent developments in astronomy implement machine learning (ML) algorithms for regression and classification in asteroid identification \cite{Sharma2022}. Our collection offers extensive asteroid data for study, backed by NASA's Jet Propulsion Laboratory \cite{BRM2023}.
Data preprocessing techniques such as feature extraction, dummy variable creation, and standardization are applied to improve model performance \cite{Singh2021b}. Predicting asteroid hazards keeps an eye out for possible dangers to Earth \cite{Ranaweera2022}. Through analyzing massive datasets and the simulation of impact scenarios, machine learning approaches increase forecasting accuracy \cite{Bacu2023} and offer crucial insights for mitigation efforts \cite{Kumar2021}. The use of AI in space science improves real-time risk assessment and forecast accuracy \cite{Shamneesh2020}. The necessity of proactive hazard prediction is underscored by past asteroid events, like the 2013 Chelyabinsk impacts and the 1908 Tunguska impacts \cite{NASA_Tunguska} \cite{Wikipedia_Chelyabinsk}.

\begin{figure}[htbp]
\centering
\includegraphics[width=6.0in]{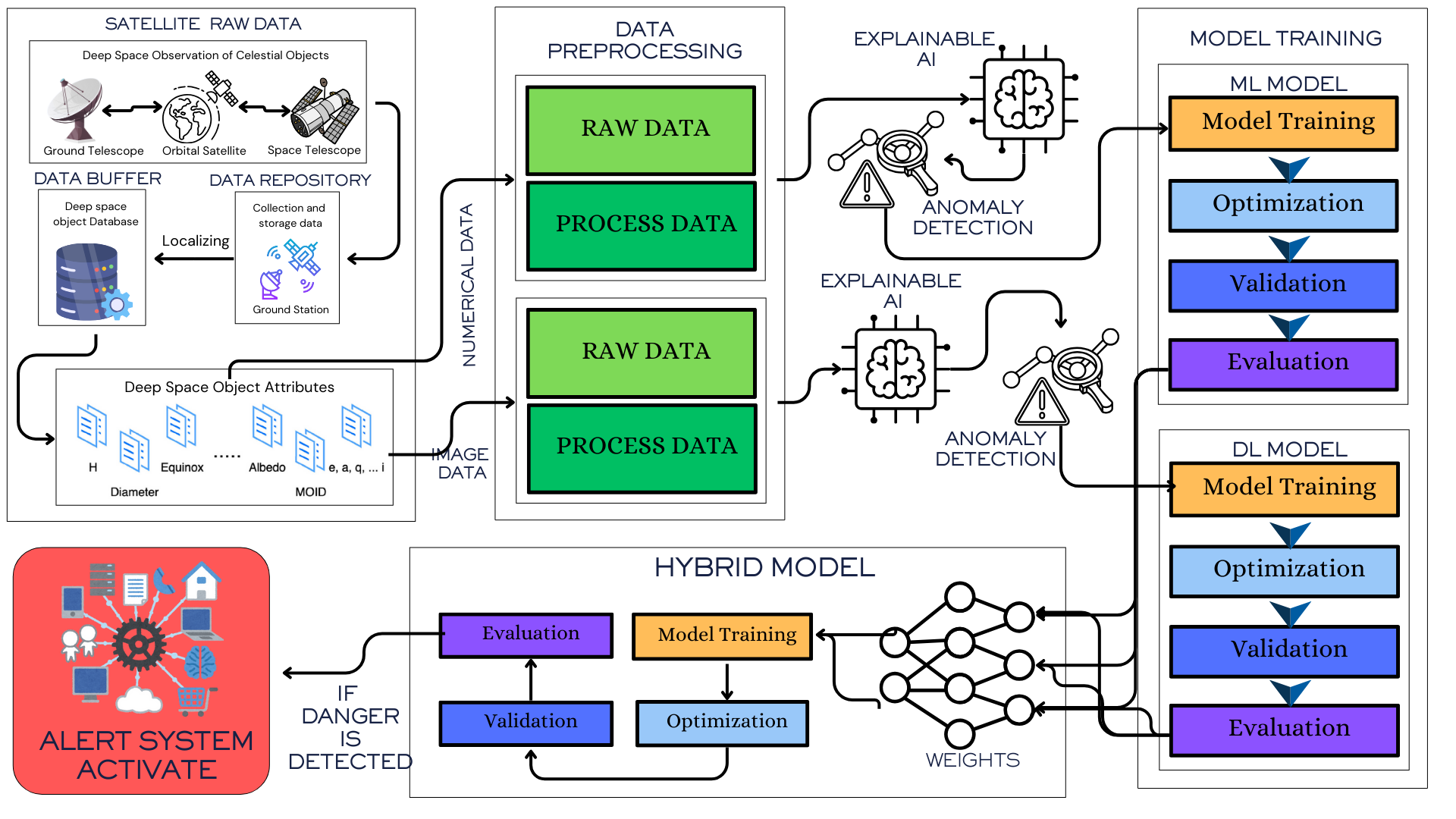}
\caption{Workflow diagram of the hazardous prediction with XAI and Anomaly detection}
\label{fig1}
\end{figure}
Numerous studies investigate different approaches to classifying asteroids. Random Forest was the top-performing model in a research that used JPL data \cite{BRM2023}. The PHAC hybrid classifier, which was shown in another study, achieved 90\% accuracy and 99\% precision \cite{Ranaweera2022}. High accuracy in asteroid classification has also been shown by deep learning models such as Inception, ResNet, and Xception \cite{Bacu2023}.
This study discuss anomaly detection and explainable AI to assess ML and DL models for asteroid categorization. To find the best-performing method, two datasets are utilize to test three DL models and seven ML algorithms. This application evaluates if deep learning works better than conventional ML models, as it has demonstrated considerable promise in classification tasks. A thorough assessment of model performance is made possible by the apply of several datasets. The accuracy of our suggested model in detecting potentially dangerous asteroids is 98.94\%. Figure 1 depicts the process of hazardous prediction, incorporating Explainable AI (XAI) for model interpretability and Anomaly Detection for predicting possible hazards. The process guarantees precise predictions while explaining decision-making.


The remaining sections of the paper are organized as
follows: The literature review is presented in Section 2, followed by the contribution of the work in Section 3, the methodology in detail and the functioning algorithm in Section 4, the testing and validation of the work along with detailed result discussion is in Section 5, and the conclusion is in Section 6.
\section{RELATED PRIOR WORKS}
Numerous studies conducted in the last few decades have improved machine learning-based classification, asteroid dynamics, and Near-Earth Object (NEO) detection. A neural network-based Hazardous Object Identifier (HOI) with 95.25\% accuracy in identifying simulated hazardous objects was created by Hefele et al. \cite{Hefele2020}. Pasko \cite{Pasko2018} used machine learning to forecast the orbital parameters of unidentified PHAs. In their evaluation of many machine learning models for classifying asteroids, Yildirim and Cigizoglu \cite{Anish2020} found that Random Forest and XGBoost had the best accuracy, while Naive Bayes had the worst, at 80.70
Si \cite{Yidirim2002} discovered that GRNN was better than MLP for estimating hydrological data. A CNN-based classifier for space objects was implemented by Linares and Roberto \cite{Linares2016} with an accuracy of 99.6\%. Asteroid diameters were predicted by Basu \cite{Basu2019} using ANN and MLP, while space debris families were analyzed by Carruba et al. \cite{Carruba2022} using ML in asteroid dynamics. For effective predictions, Naoya et al. \cite{Ozaki2022} created an ML-based asteroid flyby trajectory model.
Developed decades ago \cite{Rosenblatt1958}, machine learning has become more popular as a result of massive datasets and advancements in computing capacity. Unsupervised learning finds patterns on its own, whereas supervised learning employs labeled data for predictive analytics. CNNs are commonly employed in medical imaging through transfer learning \cite{Pardamean2018}\cite{Badza2020}, image processing, and quality assessment \cite{Varga2022} in astronomy. Large datasets are produced by telescopes, hence automated machine learning-based classification is crucial \cite{Gorgan2019}. Deep neural networks have become essential in astronomy research \cite{Takahashi2023}, made possible by improvements in hardware \cite{Sharma2016} and improved machine learning software \cite{Alam2020}.
\begin{table}[htbp]
\centering
\caption{Summary of Related Works on Asteroid Classification using ML and DL Algorithms}
\begin{tabular}
{|c|p{5cm}|p{4cm}|p{2cm}|}
\hline
\textbf{Reference} & \textbf{Objectives} & \textbf{Algorithm} & \textbf{Result} \\ \hline
\cite{Carruba2019} & The authors identified six new families and 13 new clumps that appear consistent and homogeneous regarding physical and taxonomic properties. & Supervised-learning hierarchical clustering algorithms & 89.5\% accuracy score \\ \hline
\cite{McIntyre2019} & Identifying new asteroid families and rubble pile-type rock clumps. & The k-Nearest Neighbor algorithm was combined with the Bus DeMeo taxonomic classification & 85\% average accuracy \\ \hline
\cite{Chhibber2022} & The project analyzes NASA data on Potentially Hazardous Objects (PHOs), such as asteroids and comets, to predict their threat to Earth. & Two-Class Decision Forest algorithm & 95\% ROC-AUC score \\ \hline
\cite{Pasko2018} & Predictions over the combinations of orbital parameters for yet undiscovered and potentially hazardous asteroids & Support Vector Machine algorithm with the kernel RBF. & 90\% accuracy \\ \hline
\cite{Smirnov2017} & Identified main belt asteroids, which are three body resonant. It is identified as 404 new asteroids. & KNN, Gradient Boosting, Decision Trees, Logistic Regression & 96.7\% average accuracy \\ \hline
\cite{Cowan2023} & Identifies which tracklets are entirely or partially inside the area of interest & CNN and YOLO algorithm & 97.67\% \& 90.97\% accuracy \\ \hline
\cite{Ranaweera2022} & The study explores using deep neural networks to identify patterns in NEA orbital elements and classify new asteroids as hazardous or non-hazardous efficiently. & SVM models with linear kernels, KNN, and RBF kernel & 92.02\% accuracy score \\ \hline
\cite{Bahel2021} & The study suggests using supervised machine learning to determine if an asteroid meets certain criteria for being dangerous or not & Random Forest Classifier, K-Nearest Neighbor Classifier, Decision Tree Classifier, and Logistic Regression. & 96.56\% f1 score and 96.02\% accuracy score \\ \hline
\end{tabular}

\label{tab:asteroid_summary}
\end{table}
Figure 2 illustrates the data distribution of the asteroid dataset, highlighting the variation in the important features. This plot aids in understanding feature importance and identifying possible imbalances in the dataset. Figure 3 displays sample asteroid images, providing visual insights into the dataset implemented for hazardous prediction.
\section{Contribution of the work}
\subsection{Problems Addressed in the Current Work}
Most current asteroid hazard prediction research concentrates on either ML or DL, frequently using constrained algorithms. Although quantum computing is explored in several publications, integrated ML-DL comparisons are absent. Explainable AI (XAI) for interpretability and anomaly identification for odd behaviors has not been discussed in any previous work. Accuracy is also impacted by poor data preprocessing, and most research examines textual or picture data independently rather than combining the two. Additionally, existing technologies lack the capability to provide real-time early warnings, thereby limiting proactive disaster prevention efforts.
 
\subsection{Solution Proposed}
Three DL models and seven ML methods are implemented in this work to combine ML and DL for better asteroid hazard prediction. High-performance ML and DL approaches are combined in a hybrid model to improve textual and image data analysis, while advanced preprocessing improves feature extraction. Model transparency is guaranteed by Explainable AI (XAI), and anomaly detection improves robustness by spotting odd asteroid behaviors. A real-time alert system facilitates early warnings to support prompt risk mitigation, proactive decision-making, and planetary defense initiatives.
\subsection{Novelty of the proposed Work}
The proposed work presents a hybrid method for predicting asteroidal hazards that combines ML, DL, XAI, and anomaly detection. A more accurate analysis is ensured by integrating textual and image-based data. While XAI increases transparency, advanced preprocessing improves predictive potential. Disaster preparedness is aided by a real-time alert system, which enhances early warnings. This all-encompassing strategy improves the forecast of asteroidal hazards, supports space safety and planetary defense, and offers a solid foundation for proactive risk reduction.
\section{METHODOLOGY}
\subsection{Dataset Acquisition}
We implement two essential datasets for predicting asteroid hazards: textual and image data. The textual dataset from the JPL Small-Body Database and NASA's NEO Database includes comprehensive details on over 30,000 asteroids, including hazard indications, physical characteristics, and orbital parameters, allowing for accurate numerical analysis. The 50,000+ asteroid photos in the picture dataset, which is gathered from NASA Sky Surveys and telescope archives, help with visual classification and composition research. For deep learning models, sophisticated preprocessing methods like feature extraction, augmentation, and noise reduction enhance the dataset's quality. This research ensures a more thorough approach to asteroid categorization and hazard assessment by merging both datasets and improving forecast accuracy by fusing numerical insights with visual patterns.
\begin{figure}[htbp]
\includegraphics[width=6.0in]{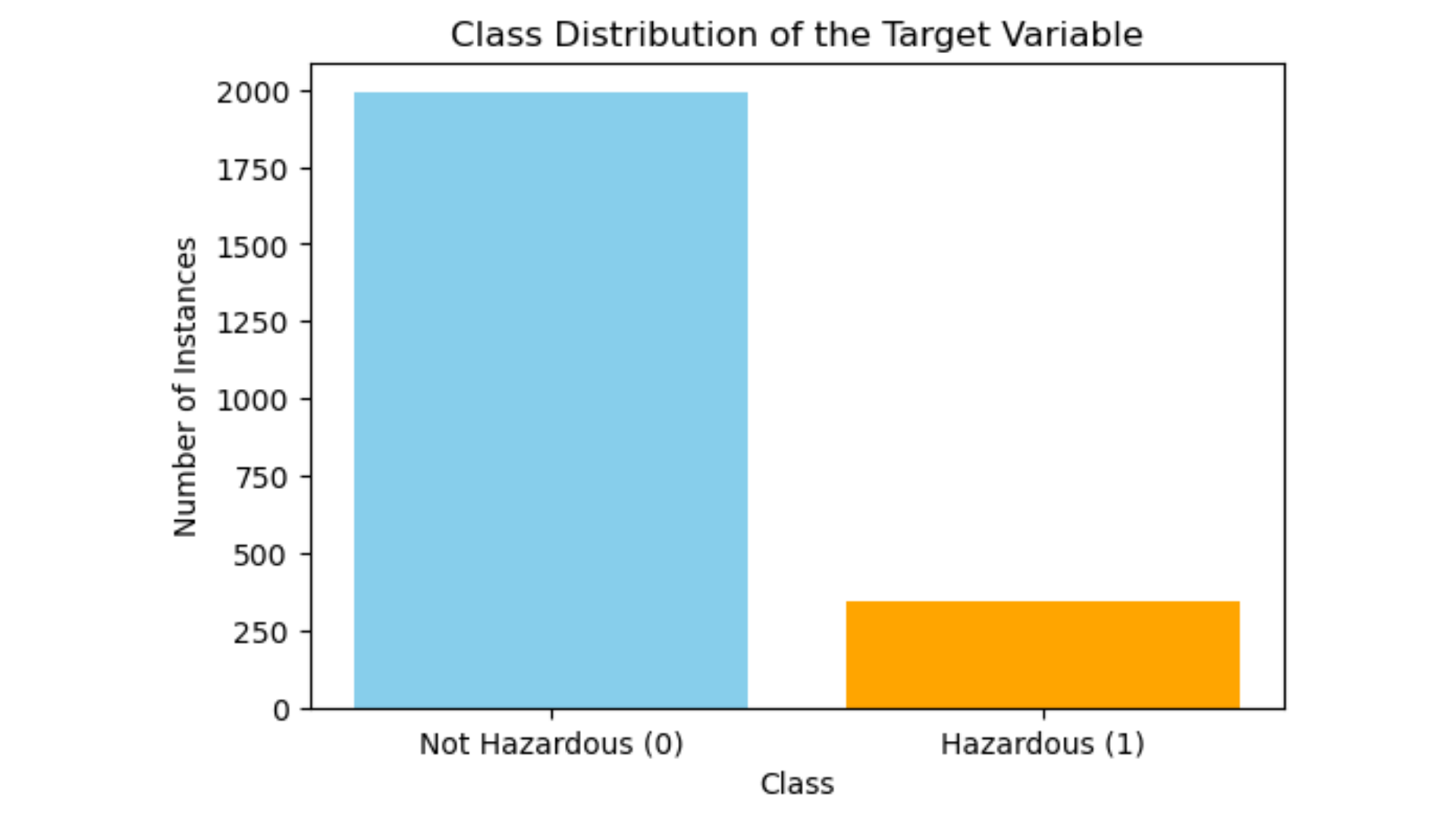}
\caption{Data Distribution of Asteroid dataset}
\label{fig_2}
\end{figure}
\begin{figure}[htbp]
\includegraphics[width=6.0in]{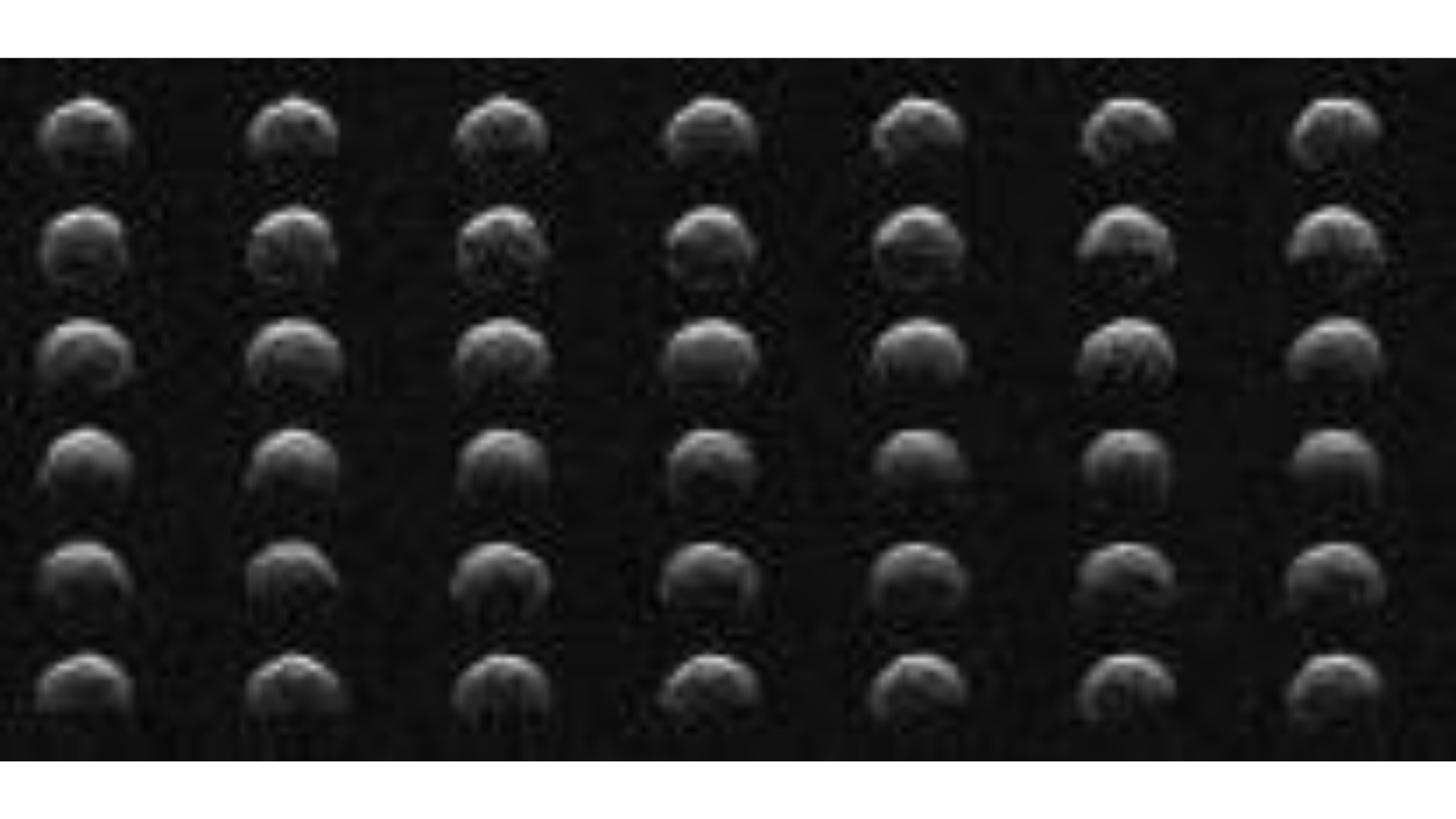}
\caption{Sample image data of Asteroid }
\label{fig_3}
\end{figure}
\subsection{Advanced Data Preprocessing Techniques}
Both textual and picture datasets are subjected to sophisticated preprocessing procedures to improve the precision and effectiveness of asteroid hazard prediction and classification. When preprocessing is done correctly, the data is clean, organized, and ready for machine learning (ML) and deep learning (DL) models.
\begin{figure}[htbp]
\includegraphics[width=6.0in]{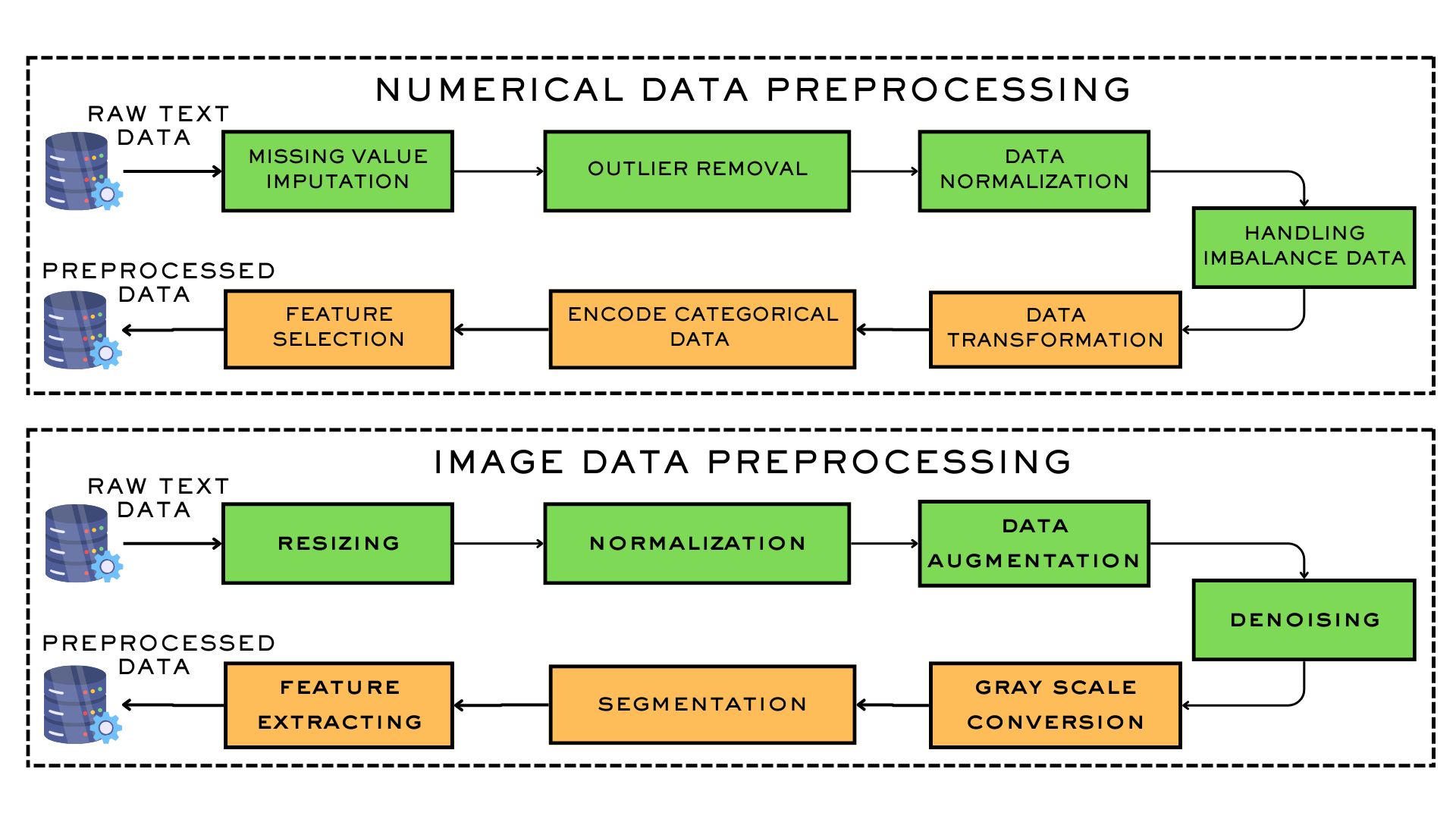}
\caption{Data Preprocessing on Numerical and Image data }
\label{fig_4}
\end{figure}
\subsubsection{Textual Data Preprocessing}
Asteroid characteristics, including orbital parameters, velocity, diameter, and hazard classification labels, are included in the textual information. K-Nearest Neighbors (KNN) and mean/median imputations are implemented to fill in missing data since real-world datasets frequently contain missing values, guaranteeing completeness. To maintain data integrity, duplicate records, and inconsistent values are detected and removed. Isolation Forest, a technique for anomaly detection that isolates rare or extreme values, is employed to detect and remove outliers in critical parameters such as eccentricity and inclination.
By guaranteeing that all features have uniform ranges, Min-Max Scaling or Standardization is need to normalize numerical values, enhancing model stability. Recursive Feature Elimination (RFE) is one feature selection strategy which is required to remove redundant information while keeping only the most pertinent asteroid features. Principal Component Analysis (PCA), which reduces noise and preserves only essential components, is utilized to improve computing efficiency and produce quicker and more accurate predictions.

\subsection{Image Data Preprocessing}
The image dataset comprises telescope-captured photographs of asteroids, which may have distortions, noise, and different lighting conditions. Gaussian and median filtering are applied to smooth the images and eliminate noise while maintaining key structures to increase clarity. By improving image contrast, adaptive histogram equalization makes subtle asteroid features more straightforward to see. 
Image augmentation techniques, including rotation, flipping, brightness modification, contrast enhancement, and zooming, are required to boost dataset variety and decrease overfitting in deep learning models, guaranteeing that the model performs well when applied to unknown data. Otsu's Thresholding and Canny Edge Detection segment asteroid sections for feature extraction help the model focus on essential parts.
Before being fed into CNN-based classification models, all images are given a consistent scale using Min-Max Normalization, which standardizes pixel intensity values between 0 and 1. To address class imbalance concerns and prevent the model from becoming biased toward non-hazardous asteroids, data balancing strategies like the Synthetic Minority Over-sampling Technique (SMOTE) are also considered. 
By significantly improving textual and image data quality, these preprocessing processes increase the model's robustness, interpretability, and accuracy in hazard prediction and asteroid categorization. 
Figure 4 shows data preprocessing for image and numerical data to make them consistent and of good quality. The process involves normalization, feature scaling, and enhancement to enhance model performance.
\begin{figure}[htbp]
\includegraphics[width=6.0in]{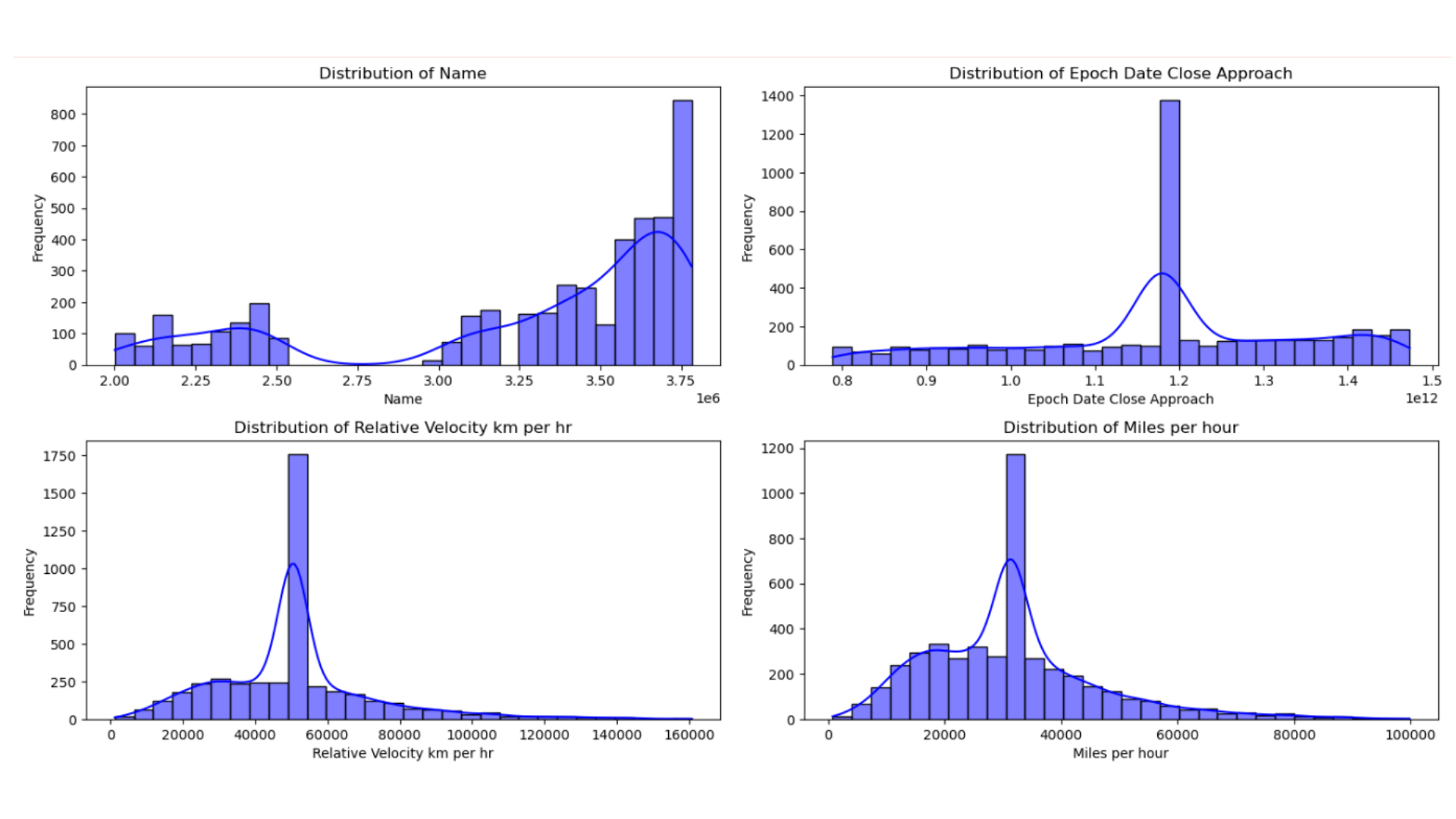}
\caption{KDE distribution before preprocessing}
\label{fig_5}
\end{figure}
\begin{figure}[htbp]
\includegraphics[width=6.0in]{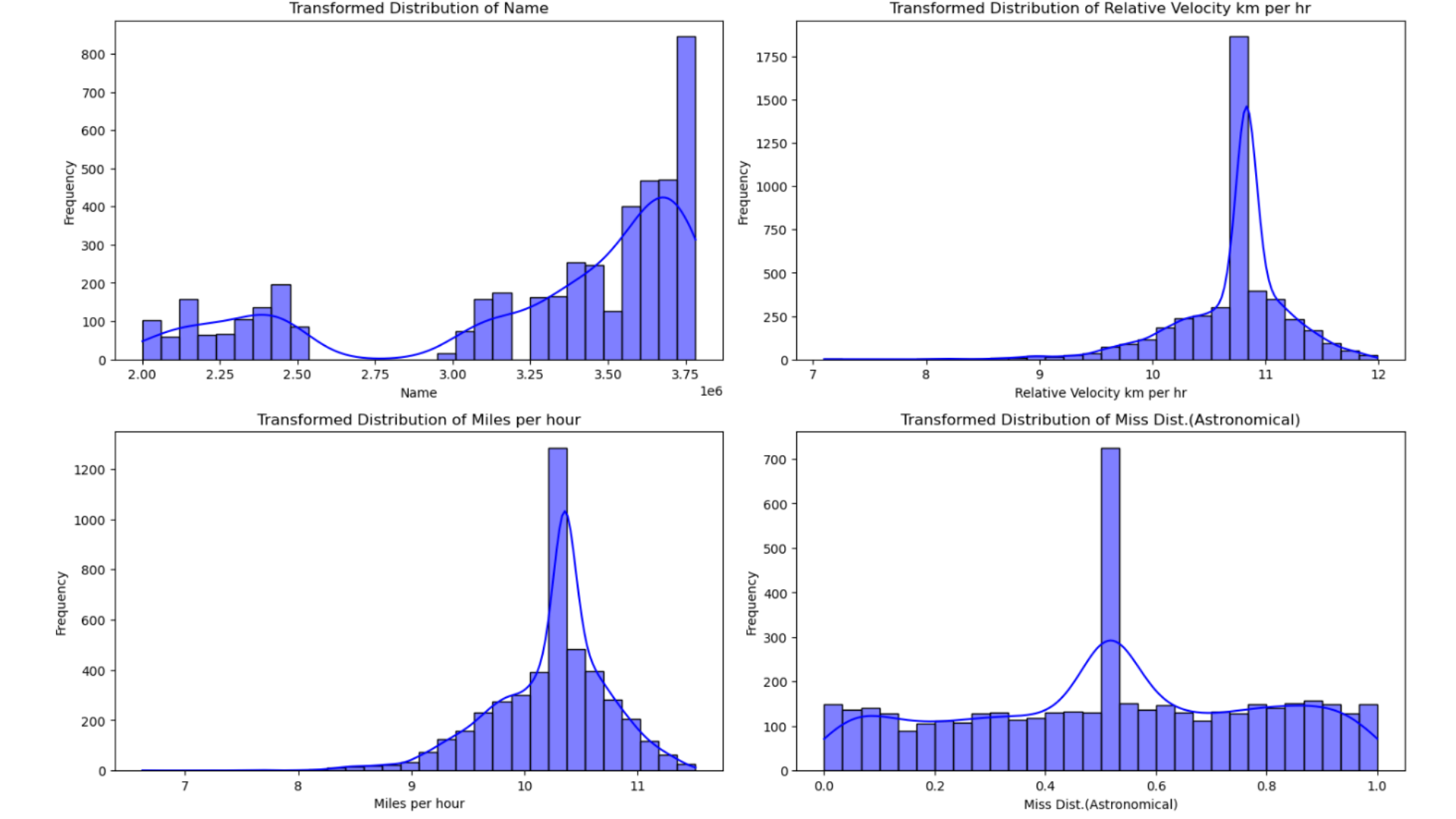}
\caption{KDE distribution after preprocessing }
\label{fig_6}
\end{figure}
\subsection{Multi-Model Approach for Asteroid Classification and Hazard Prediction}
A multi-model method that combines machine learning (ML) and deep learning (DL) ensures high accuracy in asteroid hazard classification and prediction. While DL models handle data based on images, ML models examine numerical asteroid properties. The top-performing ML and DL models are integrated into a hybrid architecture to improve performance and dependability.
\begin{table}[htbp]
    \centering
    \caption{CNN model architecture for Asteroid hazardous prediction}
    \begin{tabular}{|l|c|c|l|}
        \hline
        \textbf{Layer (Type)} & \textbf{Output Shape} & \textbf{Param \#} & \textbf{Connected to} \\
        \hline
        input\_layer (InputLayer) & (None, 262144, 3) & 0 & - \\
        dense (Dense) & (None, 262144, 64) & 256 & input\_layer[0][0] \\
        add (Add) & (None, 262144, 64) & 0 & dense[0][0] \\
        layer\_normalization (LayerNorm) & (None, 262144, 64) & 128 & add[0][0] \\
        multi\_head\_attention (MHA) & (None, 262144, 64) & 99,520 & layer\_normalization[0][0]\\ &&&layer\_normalization[0][0] \\
        add\_1 (Add) & (None, 262144, 64) & 0 & multi\_head\_attention[0][0],\\&&& add[0][0] \\
        layer\_normalization\_1 (LayerNorm) & (None, 262144, 64) & 128 & add\_1[0][0] \\
        dense\_1 (Dense) & (None, 262144, 128) & 8,320 & layer\_normalization\_1[0][0] \\
        dropout\_1 (Dropout) & (None, 262144, 128) & 0 & dense\_1[0][0] \\
        dense\_2 (Dense) & (None, 262144, 64) & 8,256 & dropout\_1[0][0] \\
        dropout\_2 (Dropout) & (None, 262144, 64) & 0 & dense\_2[0][0] \\
        add\_2 (Add) & (None, 262144, 64) & 0 & dropout\_2[0][0],\\&&& add\_1[0][0] \\
        layer\_normalization\_2 (LayerNorm) & (None, 262144, 64) & 128 & add\_2[0][0] \\
        multi\_head\_attention\_1 (MHA) & (None, 262144, 64) & 99,520 & layer\_normalization\_2[0][0],\\&&& layer\_normalization\_2[0][0] \\
        add\_3 (Add) & (None, 262144, 64) & 0 & multi\_head\_attention\_1[0][0],\\&&& add\_2[0][0] \\
        layer\_normalization\_3 (LayerNorm) & (None, 262144, 64) & 128 & add\_3[0][0] \\
        dense\_3 (Dense) & (None, 262144, 128) & 8,320 & layer\_normalization\_3[0][0] \\
        dropout\_4 (Dropout) & (None, 262144, 128) & 0 & dense\_3[0][0] \\
        dense\_4 (Dense) & (None, 262144, 64) & 8,256 & dropout\_4[0][0] \\
        dropout\_5 (Dropout) & (None, 262144, 64) & 0 & dense\_4[0][0] \\
        add\_4 (Add) & (None, 262144, 64) & 0 & dropout\_5[0][0], add\_3[0][0] \\
        layer\_normalization\_4 (LayerNorm) & (None, 262144, 64) & 128 & add\_4[0][0] \\
        \hline
    \end{tabular}
    
    \label{tab:nn_layers}
\end{table}

The table \ref{tab:nn_layers} describes the architecture of a neural network, including layer types, output shapes, parameter numbers, and connectivity. It includes dense layers for feature transformation, multi-head attention (MHA) for dependency capture, layer normalization for stability, and dropout for regularization. Add layers introduce residual connections, improving gradient flow and preventing vanishing gradients. The input shape (None, 262144, 3) progressively improves through sequential transformations, maximizing feature extraction and predictive accuracy.

Seven machine learning methods—Logistic Regression, Decision Tree, Random Forest, XGBoost, Naive Bayes, SVM, and KNN—are implemented for numerical data analysis. These algorithms all support reliable feature learning and classification. Three DL models—CNN, EfficientNetB0, and VGG16—are implemented for image-based analysis to identify significant visual patterns in asteroid photos. Combining the best features of both approaches, the hybrid methodology guarantees accurate classification and increased hazard prediction accuracy.
Figures 5 and 6 compare the KDE distribution before and after preprocessing, highlighting changes in data distribution.Figure 7 presents a heatmap plot showing feature relationships.
\begin{figure}[htbp]
\includegraphics[width=6.0in]{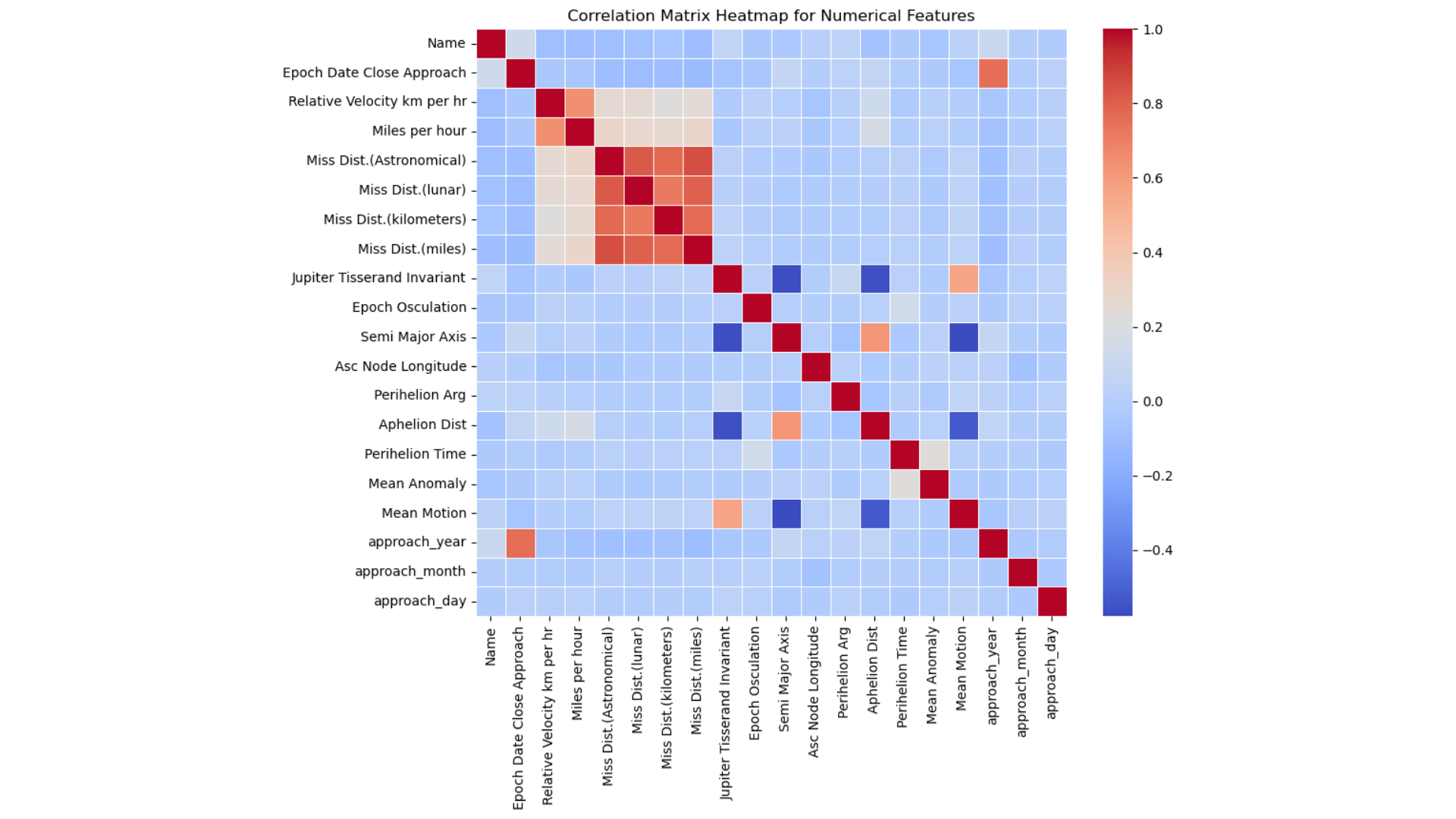}
\caption{Heatmap plot of relation between features}
\label{fig_7}
\end{figure}
\begin{algorithm}
\caption{Hybrid ML-DL Model for Asteroid Classification and Hazard Prediction}
\begin{tabbing}
\hspace*{2em} \= \hspace*{2em} \= \hspace*{2em} \= \hspace*{2em} \= \kill
\textbf{Inputs:} Preprocessed textual and image datasets of asteroids\\
\textbf{Output:} Classification of asteroids as Hazardous (1) / Non-Hazardous (0)\\

\textbf{Procedure} Hybrid\_Model(D)\\
\> $DF_{text}, DF_{image} \gets Preprocess(D)$\\
\> $train_{text}, test_{text} \gets train\_test\_split(DF_{text}, test\_size=0.2)$\\
\> $train_{image}, test_{image} \gets train\_test\_split(DF_{image}, test\_size=0.2)$\\

\> \textbf{for} model in \{LogisticRegression(), DecisionTree(), RandomForest(), SVM(), XGBoost(), NaïveBayes(), KNN()\}\\ \>\textbf{do}\\
\> \> $M_1 \gets Train(model, train_{text})$\\
\> \> $p_1 \gets M_1.predict(test_{text})$\\
\> \> $evaluate\_performance(test_{text}, p_1)$\\
\> \textbf{end for}\\

\> \textbf{for} model in \{CNN(), EfficientNetB0(), VGG16()\} \\\>\textbf{do}\\
\> \> $M_2 \gets Train(model, train_{image})$\\
\> \> $p_2 \gets M_2.predict(test_{image})$\\
\> \> $evaluate\_performance(test_{image}, p_2)$\\
\> \textbf{end for}\\

\> $H_1 \gets$ Select best-performing ML model from $\{M_1\}$\\
\> $H_2 \gets$ Select best-performing DL model from $\{M_2\}$\\
\> $Hybrid\_Model \gets Weighted\_Ensemble(H_1, H_2)$\\
\> $Final\_Prediction \gets Hybrid\_Model.predict(D)$\\

\> \textbf{if} $Final\_Prediction == 1$\\ \> \textbf{then}\\
\> \> Trigger\_Hazard\_Alert()\\
\> \textbf{end if}\\

\textbf{End Procedure}
\end{tabbing}
\end{algorithm}

\subsection{Rationale for Algorithm Selection in Asteroid Prediction}
The proper machine learning (ML) and deep learning (DL) models must applied for precise asteroid danger classification and prediction. Structured numerical data, such as asteroid orbital characteristics, velocity, size, and distance from Earth, are analyzed using machine learning algorithms. Because of its effectiveness in binary classification, logistic regression is the best method for determining whether an asteroid poses a threat. Random Forest improves accuracy and decreases overfitting, whereas Decision Trees and Random Forests efficiently record feature interactions. XGBoost needs gradient boosting to enhance prediction performance further, and Naïve Bayes uses feature distributions to classify data probabilistically. K-Nearest Neighbors (KNN) identifies asteroids by comparing them with existing data points, while Support Vector Machine (SVM) offers robust classification with distinct decision limits.
Deep learning algorithms are required for image-based asteroid classification to examine visual patterns that suggest possible dangers. Convolutional Neural Networks (CNNs) are very good at classifying images because it extract essential elements like shape, texture, and composition. Because of its optimized structure, EfficientNetB0 is favored because it offers excellent accuracy with fewer parameters, improving computational efficiency. In addition, complex asteroid structures are captured, and detailed picture categorization is ensured using VGG16, a deep CNN with many convolutional layers. By combining numerical and visual analysis for improved accuracy and risk assessment, the system guarantees a thorough and trustworthy approach to asteroid hazard prediction by integrating these ML and DL models.

\begin{figure*}[htbp]
\centering
\includegraphics[width=6.0in]{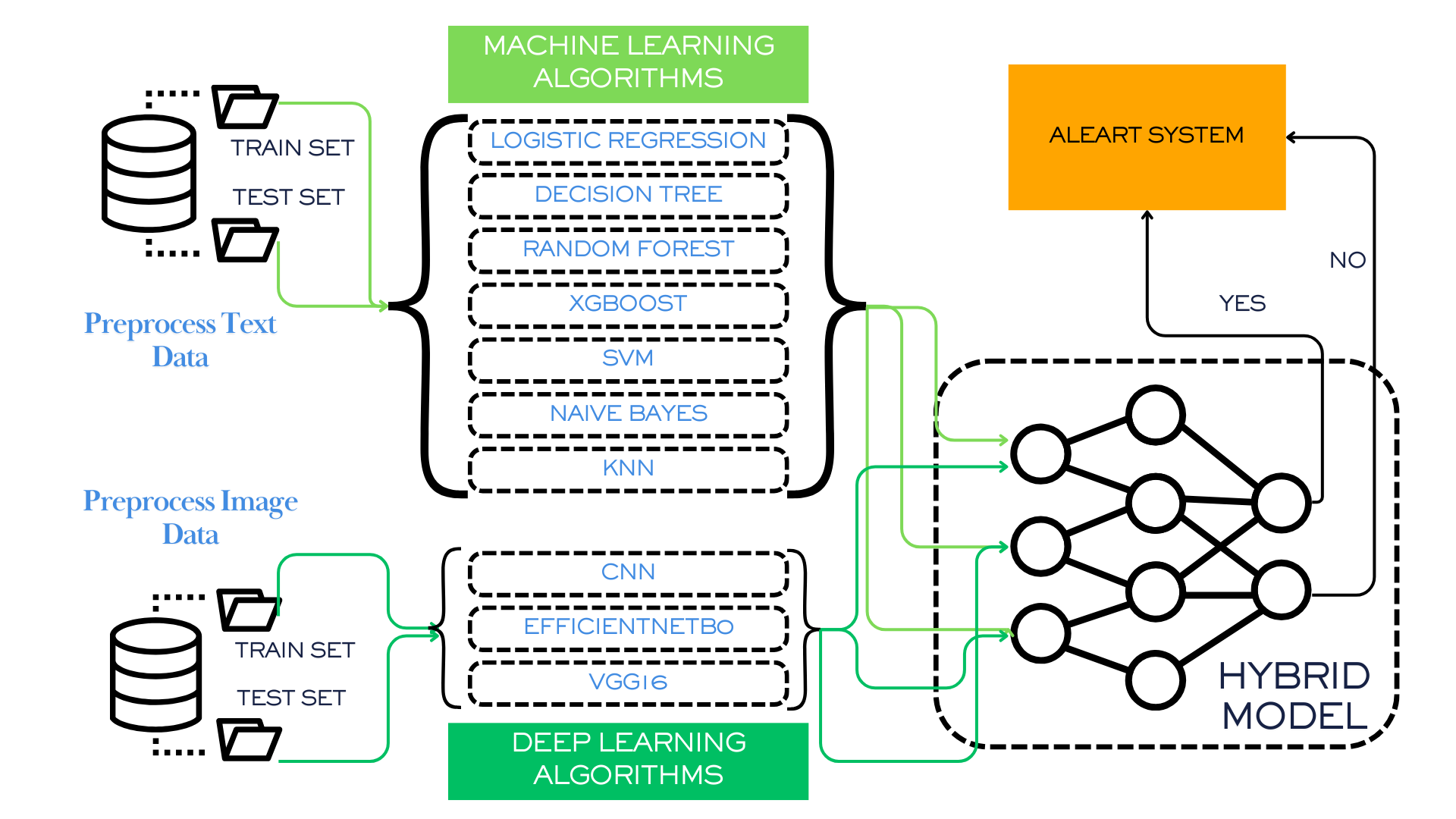}
\caption{Hybrid Algorithm specifically design for Asteroid Hazardous Prediction}
\label{fig_8}
\end{figure*}
\subsection{Proposed Hybrid Algorithm for Asteroid Classification and Hazard Prediction}
The suggested hybrid approach enhances asteroid categorization and hazard prediction by combining the top-performing deep learning (DL) and machine learning (ML) models. XGBoost and Random Forest are employed for textual data analysis, while EfficientNetB0 and VGG16 classify asteroid photos. A weighted ensemble approach yields the final prediction by optimizing model contributions according to validation performance.

This hybrid technique improves accuracy and resilience by utilizing both textual and visual data. The method guarantees a more dependable classification system by combining ML and DL, allowing for the early detection of dangerous asteroids. The model's effectiveness aids planetary defense initiatives by offering vital information for proactive risk assessment and disaster avoidance tactics. 
Figure 8 represents a hybrid algorithm for asteroid danger prediction, consisting of numerical and image data analysis. It employs SHAP XAI and anomaly detection methods integrated with deep learning to improve predictability and explainability. Algorithm 1 represents the proposed hybrid algorithm.
\subsection{Explainable AI (XAI) and Anomaly Detection in Asteroid Classification}
\subsubsection{Explainable AI (XAI)}
XAI is essential in classifying asteroids to guarantee the interpretability and transparency of model decisions. It is challenging to comprehend why an asteroid is categorized as dangerous or non-hazardous, as traditional ML and DL models operate as "black boxes." XAI approaches like SHAP (SHapley Additive exPlanations) and LIME (Local Interpretable Model-agnostic Explanations) are used to assess feature relevance and offer insights into the model's decision-making process. CNN-based algorithms use Grad-CAM (Gradient-weighted Class Activation Mapping) to identify critical areas in asteroid pictures that affect predictions. This enhances the assurance of the model and facilitates efficient result validation by astronomers.

\subsubsection{Anomaly Detection}
The current classification patterns are not always fit to identify the odd behavior of the asteroids, so anomaly detection is implemented. Outliers in asteroid trajectory, speed, and size are found using methods such as Isolation Forest, One-Class SVM, and Autoencoders. It aids in the early detection of possible risks, even if they do not fall under one of the established dangerous categories.
\subsection{Model Optimization}
Several methods are implemented to improve the accuracy and efficiency of the ML and DL models for asteroid classification. The optimal model parameters are found through hyperparameter tweaking (Grid Search, Random Search, Bayesian Optimization). Feature selection (RFE, PCA) lowers overfitting and boosts performance by ensuring that only the most pertinent data is used. For DL models, data augmentation (rotation, flipping, and noise addition) improves image variety. Training is stabilized, and overfitting is avoided via regularization (L1, L2, Dropout, Batch Normalization). DL models can be deployed more quickly and effectively by reducing computational complexity through model pruning and quantization. Together, these methods increase generalization, accuracy, and inference speed, guaranteeing accurate predictions of asteroidal hazards.
\subsection{Real-Time Alert System Deployment}
Advanced ML and DL models are integrated into a real-time alarm system to identify and reduce asteroid hazards. The trained model is made available as an API by linking with international space agencies, allowing for real-time tracking and categorization. 
For proactive decision-making, a user-friendly dashboard for mobile and online platforms offers real-time hazard visualization, showcasing impact zones, trajectories, and risk scores. The system uses a pre-trained Random Forest model for categorization after retrieving asteroid data from NASA's NEO API. When a dangerous asteroid is discovered, space agencies receive an automated email notice with important information, including the name, diameter, velocity, and approach date. It provides timely, data-driven notifications that improve planetary protection.
\begin{figure*}[htbp]
\centering
\includegraphics[width=6.0in]{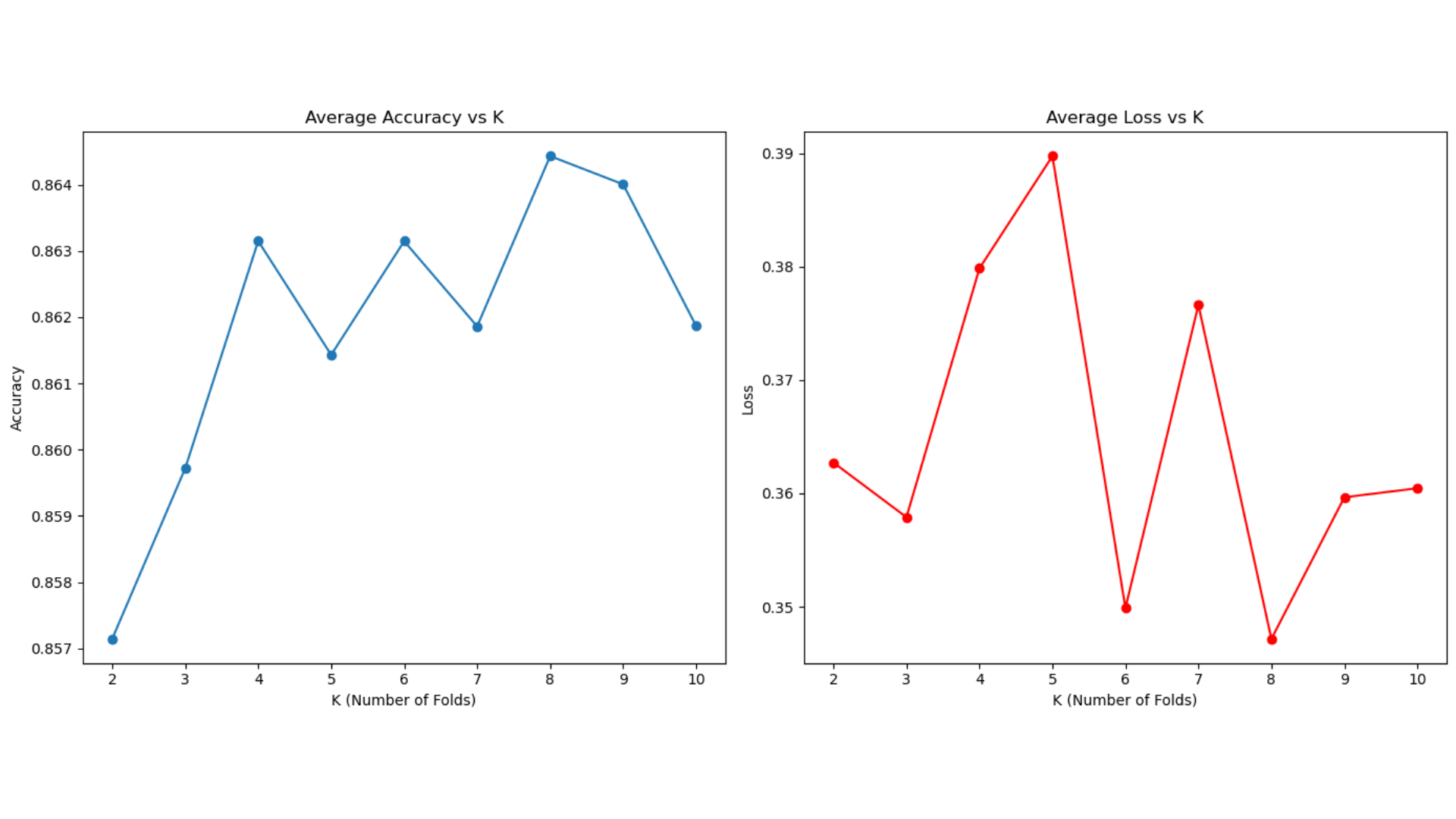}
\caption{Accuracy and Loss graph using Hybrid algorithm}
\label{fig_9}
\end{figure*}
\section{Result and Discussion}
\subsection{Experimental Setup}
The experimental setup for asteroid classification and hazard prediction is designed to handle large-scale datasets efficiently while ensuring optimal model performance.
\subsection{Hardware Configuration and Software Frameworks}
We need high-performance hardware to optimize machine learning and deep learning workflows. It includes NVIDIA A100 and RTX 3090 GPUs for accelerated computations, an Intel Xeon CPU for pre-processing and feature extraction, high-speed SSD storage for efficient data handling, and 128GB+ RAM to manage large datasets and complex operations effectively. This configuration ensures seamless execution of deep learning models and traditional machine learning tasks. The details is in Table \ref{tab:system_config}
\begin{table}[h]
    \centering
    \caption{System Configuration and Software Framework}
    \renewcommand{\arraystretch}{1.3}
    \begin{tabular}{lll}
        \hline
        \textbf{Component} & \textbf{Specification / Framework Used} & \textbf{Purpose} \\
        \hline
       GPU & NVIDIA A100,RTX3090  & Accelerated deep learning computations  \\
        
        CPU & Intel Xeon & Efficient pre-processing and traditional  \\
        && ML calculations\\
        RAM & 128GB+ & Handling large datasets and complex  \\
        &&computations\\
        Storage & High-speed SSD & Faster data access and retrieval \\
        Programming  & Python & Implementation of machine learning and \\
        Language&&deep learning models\\
        ML \& DL  & Scikit-learn, TensorFlow,  & Machine learning and deep learning  \\
       Libraries &Keras&model development\\
        XAI Tools& SHAP, LIME & Interpreting model predictions (XAI) \\
        Image Processing & OpenCV & Processing and analyzing images \\
        Data Handling & Astropy & Managing astronomical data \\
        Environment & Anaconda & Dependency management and workflow  \\
        Management&&control\\
        Database& MongoDB, Postgre & Storing and managing structured data  \\
        &SQL&efficiently\\
        API & Flask, FastAPI & Deploying APIs for model inference and  \\
        Implementation&&interaction\\
        \hline
    \end{tabular}
    
    \label{tab:system_config}
\end{table}

The proposed work is implemented in Python, utilizing robust libraries like Scikit-learn for machine learning and TensorFlow and Keras for deep learning. Explainable AI (XAI) interprets model predictions using SHAP and LIME. Astropy makes it easier to handle astronomical data, and OpenCV helps with picture processing. Anaconda is used to control the environment, guaranteeing smooth dependency management. Furthermore, MongoDB/PostgreSQL is integrated for adequate asteroid data storage, and Flask/FastAPI is utilized for API implementation. 
\begin{table}[htbp]
    \centering
    \caption{Accuracy metrics on text data with Advance Preprocessing, XAI and Anomaly Detection}
    \begin{tabular}{lcccc}
        \toprule
        \textbf{Model Name} & \textbf{Accuracy Score} & \textbf{Precision Score} & \textbf{Recall Score} & \textbf{F1 Score} \\
        \midrule
        Logistic Regression & 0.97 & 0.95 & 0.96 & 0.96 \\
        Decision Tree       & 0.95 & 0.93 & 0.94 & 0.94 \\
        Random Forest       & 0.98 & 0.97 & 0.97 & 0.97 \\
        Support Vector Machine & 0.97 & 0.96 & 0.96 & 0.96 \\
        XGBoost             & 0.99 & 0.98 & 0.98 & 0.98 \\
        Naïve Bayes         & 0.94 & 0.92 & 0.93 & 0.93 \\
        KNN                 & 0.96 & 0.94 & 0.95 & 0.95 \\
        \bottomrule
    \end{tabular}
\end{table}

\begin{table}[htbp]
    \centering
    \caption{Comparison between Accuracy score of different types of methods}
    \begin{tabular}{lccc}
        \toprule
        \textbf{Model Name} & \textbf{Accuracy Without } & \textbf{Accuracy With } & \textbf{Accuracy With XAI } \\
        &\textbf{Advanced Preprocessing}&\textbf{Advanced Preprocessing}&\textbf{and Anomaly Detection}\\
        \midrule
        Logistic Regression  & 0.87 & 0.94 & 0.97 \\
        Decision Tree        & 0.86 & 0.92 & 0.95 \\
        Random Forest        & 0.91 & 0.96 & 0.98 \\
        Support Vector Machine & 0.89 & 0.95 & 0.97 \\
        XGBoost             & 0.92 & 0.97 & 0.99 \\
        Naïve Bayes         & 0.83 & 0.91 & 0.94 \\
        KNN                 & 0.88 & 0.94 & 0.96 \\
        \bottomrule
    \end{tabular}
\end{table}
\subsection{Detail Discussion}
The analysis of several machine learning models demonstrates how enhanced preprocessing and the combination of anomaly detection and Explainable AI (XAI) majorly affect model performance. Models trained without sophisticated preprocessing initially showed modest accuracy, with Naïve Bayes performing the lowest at 83\% and XGBoost the highest at 92\% as in Table 3. Accuracy increased in all models following the application of sophisticated preprocessing techniques such as feature scaling, outlier removal, and data balance; Random Forest achieved 96\% and XGBoost 97\%, as shown in Table 4.
\begin{figure*}[htbp]
\centering
\includegraphics[width=6.0in]{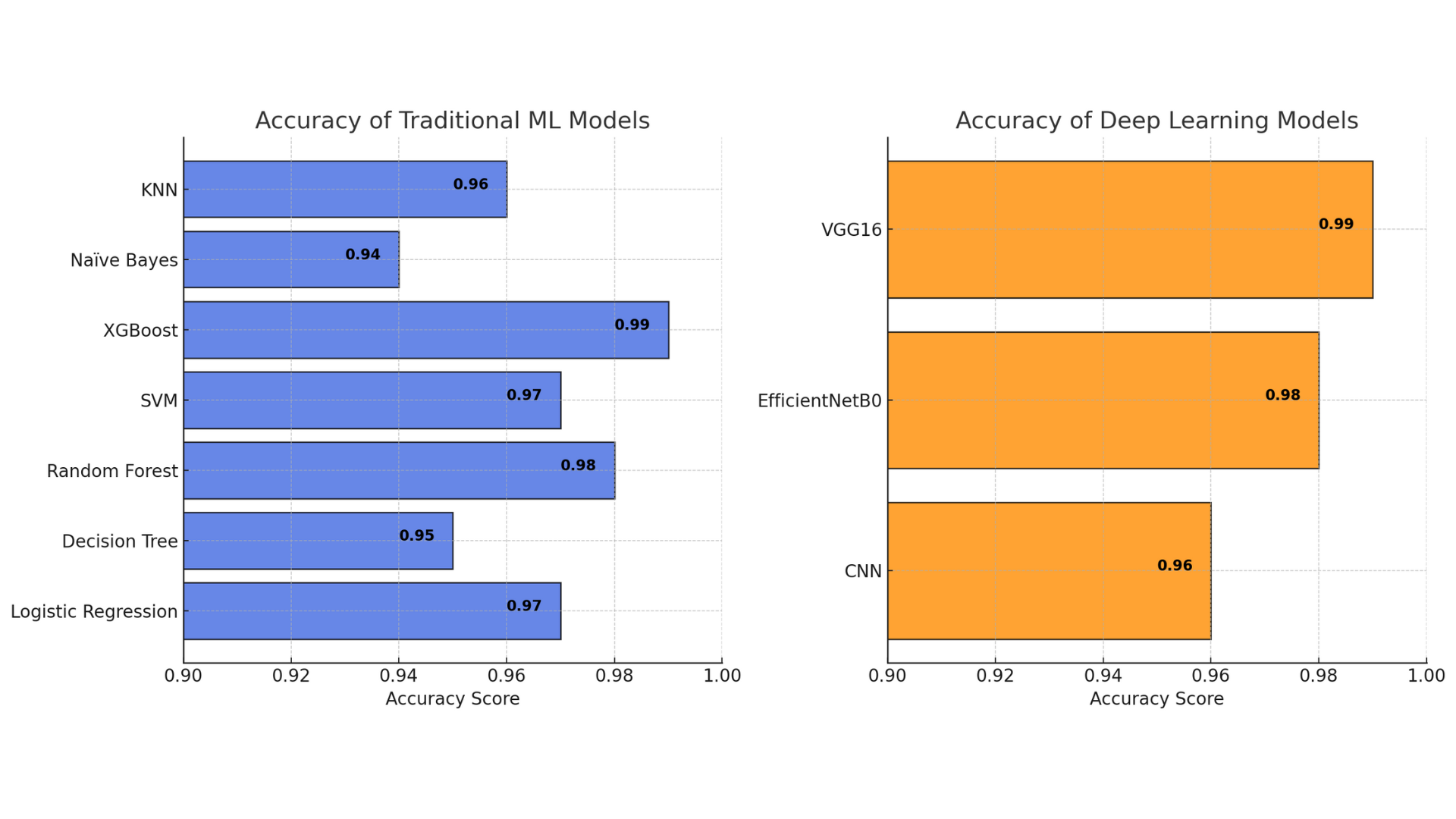}
\caption{Barplot of ML and DL Algorithms}
\label{fig_10}
\end{figure*}
\begin{figure*}[htbp]
\centering
\includegraphics[width=6.0in]{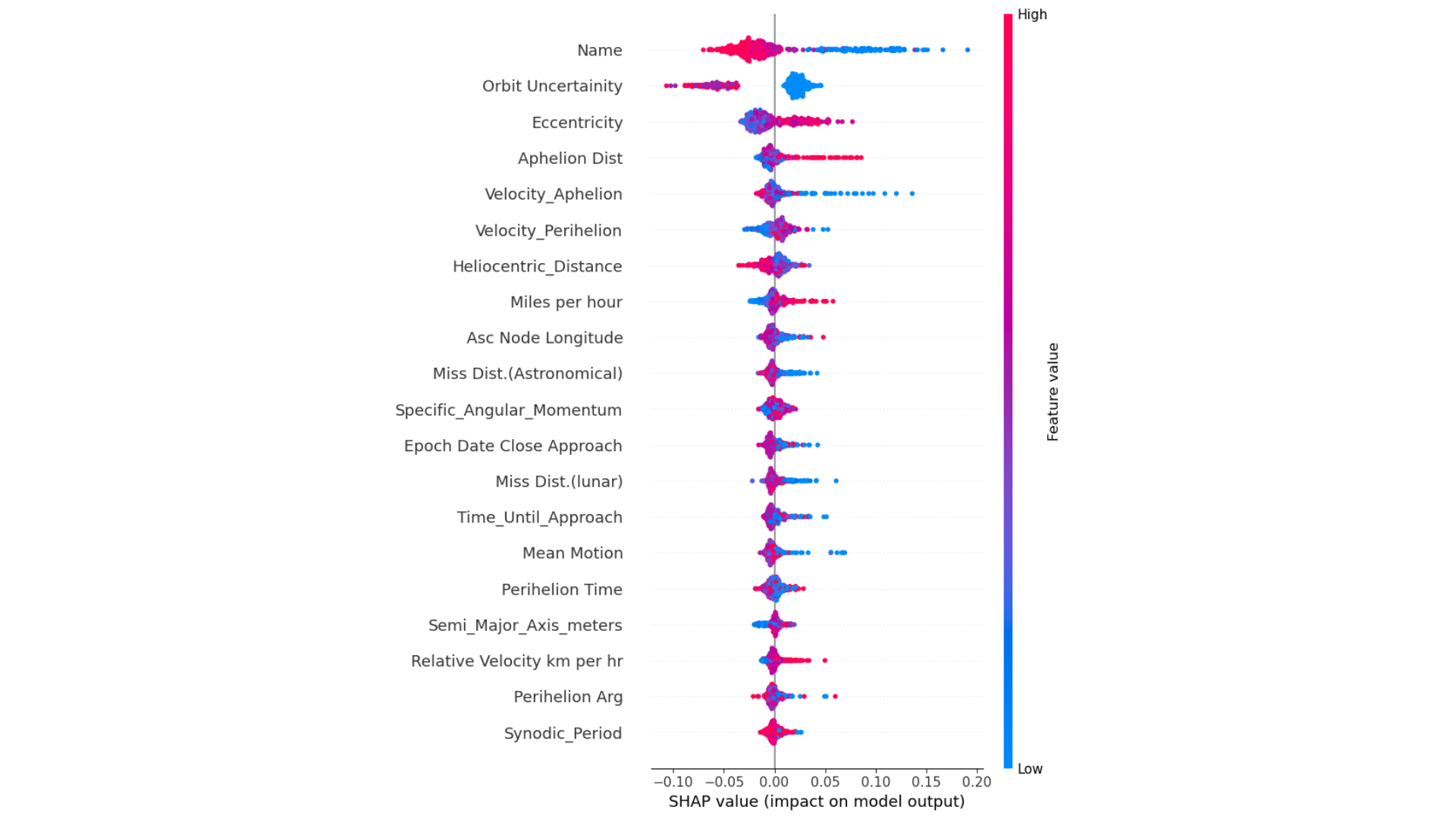}
\caption{SHAP value of feature selection}
\label{fig_12}
\end{figure*}
\begin{figure*}[htbp]
\centering
\includegraphics[width=6.0in]{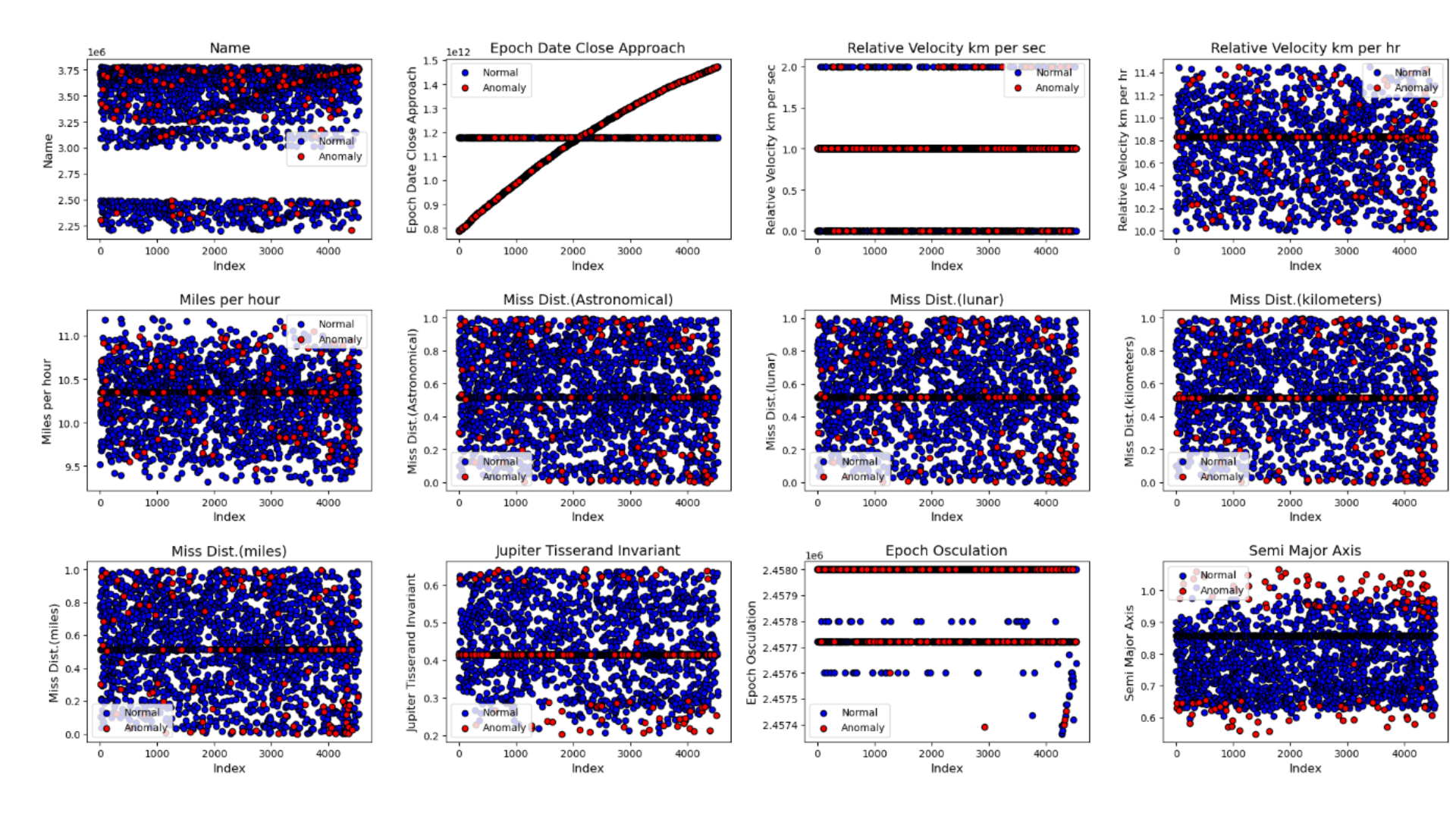}
\caption{Anomaly detection on Asteroid data}
\label{fig_13}
\end{figure*}
Additional improvements are shown when XAI and anomaly detection are added, which enhances model interpretability and more skillfully manages data anomalies. With an accuracy of 99\%, XGBoost beat all other models, while Random Forest came in second with 98\%. The Precision, Recall, and F1 scores demonstrated similar patterns, with XGBoost continuously obtaining the top scores (0.98 across all metrics). These results show that interpretability and anomaly detection methods significantly improve model performance and dependability when combined with a well-organized preprocessing pipeline. Figure 9 depicts the accuracy and loss graphs for the hybrid algorithm, showcasing the model's training performance. The trends indicate effective learning, decreasing loss, and improving epoch accuracy. Figure 10 presents a bar plot comparing ML and DL algorithm performance.

\begin{table}[htbp]
    \centering
    \caption{Performance Metrics of Deep Learning Models}
    \begin{tabular}{lcccc}
        \toprule
        \textbf{Model Name} & \textbf{Accuracy Score} & \textbf{Precision Score} & \textbf{Recall Score} & \textbf{F1 Score} \\
        \midrule
        CNN               & 0.96 & 0.94 & 0.95 & 0.95 \\
        EfficientNetB0    & 0.98 & 0.97 & 0.97 & 0.97 \\
        VGG16             & 0.98 & 0.98 & 0.98 & 0.98 \\
        \bottomrule
    \end{tabular}
\end{table}

\begin{table}[htbp]
    \centering
    \caption{Accuracy Comparison of Deep Learning Models with Different Techniques}
   \begin{tabular}{lccc}
        \toprule
        \textbf{Model Name} & \textbf{Accuracy Without } & \textbf{Accuracy With } & \textbf{Accuracy With XAI } \\
        &\textbf{Advanced Preprocessing}&\textbf{Advanced Preprocessing}&\textbf{and Anomaly Detection}\\
        \midrule
        CNN                & 0.88 & 0.94 & 0.96 \\
        EfficientNetB0     & 0.91 & 0.96 & 0.98 \\
        VGG16              & 0.92 & 0.97 & 0.98 \\
        \bottomrule
    \end{tabular}
\end{table}
\begin{table}[htbp]
    \centering
    \caption{Comparison of Related Works and Proposed Approach}
    \begin{tabular}{p{3cm} p{5cm} p{3cm} p{2cm}}
        \toprule
        \textbf{Reference} & \textbf{Objectives} & \textbf{Algorithm} & \textbf{Accuracy} \\
        \midrule
        \textbf{\cite{Gupta2023}} & Classical ML Model comparative analysis with project quantum kernel feature & DT,KNN,SVC,Naïve Bayes & 92\%  \\
        \textbf{\cite{Ranaweera2022}} & Develop a classifier to detect potentially hazardous asteroids automatically & SVM, LR, KNN, and MLP &  90\%  \\
        \textbf{\cite{Chhibber2022}} & Analyze NASA’s data on asteroids and use ML to predict hazardous approaches & DT,Boosted DT,SVM,LR & Avg 96\%  \\
        \textbf{Proposed Work} & ML on text data with advanced preprocessing, XAI, and anomaly detection  & RF, XGBoost, SVM, LR &  99\%  \\
        \textbf{} & DL on image data with advanced preprocessing, XAI to find hidden insights & CNN, EffNetB0, VGG16 &  98\%  \\
        \bottomrule
    \end{tabular}
\end{table}

According to the evaluation results, VGG16 performs better than CNN and EfficientNetB0, attaining the maximum accuracy (0.99) when using anomaly detection and XAI approaches. Additionally, EfficientNetB0 exhibits strong generalization with an accuracy of 0.98 following sophisticated processing. With an F1 score of 0.95, CNN performs well even if it is marginally less accurate, as shown in Tables 5 and 6.

The efficiency of explainability and data enhancement strategies in deep learning-based categorization is demonstrated by increased accuracy following sophisticated preparation and XAI integration.
Name (0.0368) is the most significant feature in the dataset, suggesting that specific asteroid names might have predictive value due to naming traditions based on their attributes. Closely behind, orbit uncertainty (0.0331) illustrates how crucial trajectory stability is in predicting asteroid behavior. Motion predictions are significantly impacted by eccentricity (0.0184), a measure of an orbit's departure from a complete circle. Understanding orbital trajectories requires knowledge of the aphelion distance, the furthest point from the Sun (0.0088). Finally, by affecting an asteroid's speed at its furthest orbital point, Velocity at Aphelion (0.0086) influences the impact probability. Figure 11 presents the SHAP values for feature selection, pointing out the most significant factors in predicting asteroid hazards. Figure 12 displays anomaly detection used on asteroid data, pinpointing outliers that can be potential threats. Both these methods improve model interpretability and detection rates. Table 7 compares existing and proposed work in terms of accuracy score.

The study emphasizes how Explainable AI (XAI), anomaly detection, and sophisticated preprocessing have a significant effect on how well machine learning (ML) and deep learning (DL) models perform. According to the findings, a well-organized preprocessing pipeline increases model accuracy; further improvements are seen when XAI and anomaly detection methods are incorporated. The proposed work provides the path in the domain with several pieces of information. We observe that the enhanced preprocessing plays a crucial role in improving model performance. Techniques such as feature scaling, outlier detection, and data balancing significantly improve predictive accuracy and should be prioritized before model training. Integrating XAI improves interpretability by providing insights into feature importance and increasing model clarity and reliability, primarily in critical applications like healthcare and asteroid risk prediction. Anomaly detection further boosts model reliability by identifying and mitigating data anomalies, making its incorporation into ML pipelines essential. Hybrid approaches combining traditional ML models with deep learning architectures have demonstrated superior performance, suggesting that ensemble techniques could further optimize results. Integrating preprocessing, XAI, and anomaly detection also enables scalability and generalizability, making these models applicable across various domains, including healthcare, finance, and space exploration. 

\section{Conclusion}
This research proposes a hybrid ML-DL approach for predicting hazards  and categorizing asteroids with enhanced accuracy using anomaly detection, XAI, and advanced preprocessing. The findings recommends for adding advanced preprocessing, feature selection, and interpretability features boosts the model's performance. Significantly, the highest accuracy was recorded by XGBoost at 99\%, followed by Random Forest with 98\%, and VGG16 surpassed other DL models at 98\% accuracy.
The comparison with other approaches points to the enhanced prediction ability of the suggested approach. Incorporating Explainable AI ensures that the model's decisions are understandable, which is critical in instilling confidence in AI-based hazard prediction. Anomaly detection methods further aid in discovering infrequent yet essential asteroid activities, enhancing the robustness of the classification framework.
A thorough comparison of model performances at various levels of preprocessing indicates the need for organized data-boosting methods. Progressive improvement in precision, from base models to their XAI-enhanced and anomaly-detection forms, supports the strength of hybrid methodology. The extension of the integrated framework with real-time alert management further increases this approach's value in practical scenarios, providing warnings and interpretations for planetary defense missions.

In future studies, additional astrophysical datasets will be integrated, and the framework will be optimized for large-scale real-time applications. Expanding this technique to more general space object classification problems and optimizing the hybrid model's performance will further establish its place in planetary protection and space exploration.



\bibliographystyle{unsrt}  

\end{document}